\documentclass[12pt]{iopart}

\usepackage{graphicx}
\usepackage{dcolumn}
\usepackage{amsgen}
\usepackage{amsfonts}
\usepackage{amssymb}
\usepackage{amsbsy}
\usepackage{bm}
\usepackage{color}
\usepackage{epstopdf}

\usepackage{verbatim,times,bbm}

\begin{document}

\newcommand{\beq}{\begin{equation}}
\newcommand{\eeq}{\end{equation}}
\newcommand{\barr}{\begin{eqnarray}}
\newcommand{\earr}{\end{eqnarray}}
\newcommand{\andy}[1]{ }
\newcommand{\bmsub}[1]{\mbox{\boldmath\scriptsize $#1$}}
\def\bra#1{\langle #1 |}
\def\ket#1{| #1 \rangle}
\def\sinc{\mathop{\text{sinc}}\nolimits}
\def\cV{\mathcal{V}}
\def\cH{\mathcal{H}}
\def\cT{\mathcal{T}}
\def\cP{\mathcal{P}}
\renewcommand{\Re}{\mathop{\mathrm{Re}}\nolimits}
\renewcommand{\Im}{\mathop{\mathrm{Im}}\nolimits}

\newcommand{\REV}[1]{\textrm{\color{red}#1}}
\newcommand{\BLUE}[1]{\textbf{\color{blue}#1}}
\newcommand{\GREEN}[1]{\textbf{\color{green}#1}}
\newcommand{\blu}[1]{\textbf{\color{blue}#1}}

\title[Multipartite Entanglement and Frustration]{Multipartite Entanglement and Frustration}

\author{P. Facchi $^{1,2}$, G. Florio$^{3,2}$, U. Marzolino$^{4}$, G. Parisi$^{5}$, S. Pascazio$^{3,2}$}

\address{$^1$Dipartimento di Matematica, Universit\`a di Bari, I-70125 Bari, Italy}
\address{$^2$Istituto Nazionale di Fisica Nucleare, Sezione di Bari, I-70126 Bari, Italy}
\address{$^3$Dipartimento di Fisica, Universit\`a di Bari, I-70126 Bari, Italy}
\address{$^4$Dipartimento di Fisica, Universit\`a di Trieste, and Istituto Nazionale di Fisica Nucleare, Sezione di Trieste, I-34014 Trieste, Italy}
\address{$^5$Dipartimento di Fisica, Universit\`{a} di Roma ``La Sapienza", Piazzale Aldo Moro 2, \\
Centre for Statistical Mechanics and Complexity (SMC), CNR-INFM, \\
and Istituto Nazionale di Fisica Nucleare, Sezione di Roma, 00185
Roma, Italy}

\begin{abstract}
Some features of the global entanglement of a composed quantum system can be quantified in terms of the purity of a balanced bipartition,  made up of half of its subsystems. For the given bipartition, purity can always be minimized by taking a suitable (pure) state. When many bipartitions are considered, the requirement that purity be minimal for all bipartitions can engender conflicts and frustration arises. This unearths an interesting link between frustration and multipartite entanglement, defined as the average purity over all (balanced) bipartitions. 
\end{abstract}


\maketitle

\section{Introduction}\label{sec:intro}

Frustration in humans and animals arises from unfulfilled needs. 
Freud related frustration to goal attainment and identified inhibiting conditions that hinder the realization of a given objective \cite{freud}. 
In the psychological literature one can find many diverse definitions, but roughly speaking, a situation is defined as frustrating when a physical, social, conceptual or environmental obstacle prevents the satisfaction of a desire \cite{barker}.
Interestingly, definitions of frustration have appeared even in the jurisdictional literature and appear to be related to an increased incidence of parties seeking to be excused from performance of their contractual obligations \cite{lawliterature}. There, 
``Frustration occurs whenever the law recognises that without default of either party a contractual obligation has become incapable of being performed because the circumstances in which performance is called for would render it a thing radically different from that which was undertaken by the contract.... It was not this I promised to do."  \cite{law}

In physics, this concept must be mathematized.
A paradigmatic example \cite{parisi} is that of three characters, $A, B$ and $C$, who are not good friends and do not want to share a room. However, there are only \emph{two} available rooms, so that at least two of them will have to stay together. Their needs will therefore not be fulfilled and frustration will arise. 
A schematic mathematical description of this phenomenon consists in assigning a ``coupling" constant $J_{ik}$ to each couple $(i,k)$, with $i,k=A,B,C$: $J_{ik}=+1(-1)$ if $i$ and $k$ (do not) like to share a room. Each character is then assigned a dichotomic variable $S_i=+1(-1)$ if $i$ is in the first (second) room. The key ingredient is the definition of a \emph{cost function} that quantifies the amount of ``discomfort" (unfulfilled needs) of our three characters. This can be easily done:
\begin{equation}\label{eq:hparisi}
H 
= -\frac{1}{2} \sum_{i\neq j} J_{ik}S_iS_k.
\end{equation}
The goal is to minimize this cost. In our case $J_{ik}=-1, \forall i,k$, so that 
\begin{equation}\label{eq:hparisi1}
H = S_AS_B + S_AS_C + S_BS_C .
\end{equation}
Each addendum in the summation can take only two values, $\pm1$. However, although each single addendum can be made equal to $-1$ (separate rooms, minimum cost and no discomfort for the given couple), their sum, in the best case, is $H_{\rm min}=-1$, which is larger than the sum of the three minima, $-3$. At least two characters will have to share a room and frustration arises. The situation becomes more complicated (and interesting) when more characters are involved and the coupling constants in 
(\ref{eq:hparisi}) are, e.g., statistically distributed.

The above description of frustration, in terms of a cost function, applies to a classical physical system. Interestingly, there is a frustration associated with quantum entanglement in many body systems. The study of this problem will be the object of the present investigation.

Entanglement is a very characteristic trait of quantum mechanics, that was identified at the dawn of the theory \cite{EPR,Schr1,Schr2}, has no analogue in classical physics \cite{entanglement} and came recently to be viewed 
as a resource in quantum information science \cite{entanglementrev,h4}. 
When the system is bipartite, its entanglement can be unambiguously quantified in
terms of the von Neumann entropy or the entanglement of formation
\cite{wootters,entanglement2}. Difficulties arise, however, when one endeavours to define \emph{multipartite} entanglement 
\cite{multipart1,multipart2,multipart3,multipart4,Bergou}. The main roadblock is due to the fact that states endowed with large
entanglement typically involve exponentially many coefficients and cannot be quantified with a few measures. 
This interesting feature of multipartite entanglement, already alluded to in Ref.\  \cite{MMSZ}, motivated us to look for a statistical approach \cite{statmech}, based on a characterization of entanglement that makes use of the probability density function of the entanglement of a subsystem over all (balanced) bipartitions of the total system \cite{FFP}. A state has a large multipartite entanglement if its average bipartite entanglement is large (and possibly also largely independent of the bipartition).

Maximally multipartite entangled states 
(MMES) \cite{MMES,multent} are states whose entanglement is maximal for every (balanced) bipartition. The study of MMES has brought to light the presence of frustration in the system, highlighting the complexity inherent in the phenomenon of multipartite entanglement. Frustration in MMES is due to a ``competition" among biparititions and the impossibility of fulfilling the requirement of maximal entanglement for all of them, given the quantum state \cite{MMES, statmech}. The links between entanglement and frustration were also investigated in Refs.\ \cite{frust1,frust2,frust3}.

This paper is organized as follows. We introduce notation and define
maximally bipartite and maximally multipartite entangled states in
Sec.\ \ref{sec:definitions}. 
We numerically investigate these states and 
show that multipartite entanglement is a complex phenomenon and
exhibits frustration, whose features are studied in Sec.\
\ref{sec:frustr7}. Section
\ref{sec:concl} contains our conclusions and an outlook.

\section{From bipartite to multipartite entanglement}
\label{sec:definitions}

The notion of MMES was originally introduced for qubits \cite{MMES} and then extended to continuous variable systems \cite{Adesso,FFLMP}. Here we follow \cite{FFLMP} and give a system-independent formulation.
Consider a system composed of $n$ identical (but distinguishable) subsystems.
Its Hilbert space $\mathcal{H}=\mathcal{H}_S$, with
$\mathcal{H}_S := \bigotimes_{i\in S} \mathfrak{h}_i$ and
$S=\{1,2,\dots,n\}$, is the tensor product of the Hilbert
spaces of its elementary constituents $\mathfrak{h}_i\simeq \mathfrak{h}$. Examples
range from qubits, where $\mathfrak{h}=\mathbb{C}^2$, to continuous
variables systems, where $\mathfrak{h}=L^2(\mathbb{R})$. We will
denote a bipartition of system $S$ by the pair $(A,\bar{A})$, where
$A\subset S$, $\bar{A}= S \setminus A$ and $1\leq n_A \leq
n_{\bar{A}}$, with $n_A=|A|$, the cardinality of party $A$ ($n_A+n_{\bar{A}}=n$). At the level of Hilbert spaces we get
\begin{equation}
\mathcal{H}=\mathcal{H}_A \otimes \mathcal{H}_{\bar{A}}.
\end{equation}

Let the total
system be in a pure state $|\psi\rangle \in \mathcal{H}$, which is
the only case we will consider henceforth.
The amount of entanglement between party $A$ and party $\bar A$ can be quantified, for instance,
in terms of the purity
\begin{equation}\label{eq:piAdef}
\pi_A = \tr (\rho_A^2)
\end{equation}
of the reduced density matrix of party $A$,
\begin{equation}
\rho_A = \tr_{\mathcal{H}_{\bar{A}}} (|\psi\rangle\langle\psi|) . 
\end{equation}
Purity ranges between
\begin{equation}
\label{eq:piArange}
\pi_{\mathrm{min}}^{n_A} \leq \pi_A \leq 1,
\end{equation}
where
\begin{equation}
\label{eq:piallArange}
\pi_{\mathrm{min}}^{n_A}=  (\dim \mathcal{H}_A)^{-1}=(\dim \mathfrak{h})^{-n_A},
\end{equation}
with the stipulation that $1/\infty = 0$.
The upper bound $1$ is attained by unentangled, factorized states
$|\psi\rangle=|\phi\rangle_A\otimes |\chi\rangle_{\bar{A}}$ (according to the given bipartition).  When $\dim \mathfrak{h}<\infty$, the lower bound, that
depends only on  the number of elements $n_A$ composing party $A$,
is attained by maximally bipartite entangled states, whose reduced
density matrix is a completely mixed state
\begin{equation}
\rho_A=\pi_{\mathrm{min}}^{n_A}  \mathbf{1}_{\mathcal{H}_{A}}
\end{equation}
where $\pi_{\mathrm{min}}^{n_A}$ is defined in Eq.(\ref{eq:piallArange}) and $\mathbf{1}_{\mathcal{H}_{A}}$ is the identity operator on $\mathcal{H}_{A}$.
This property is valid at fixed bipartition  $(A,\bar{A})$; we now try and extend it to more bipartitions.

Consider the average purity (``potential of multipartite entanglement") \cite{scott,MMES}
\begin{equation}
\label{eq:piave}
\pi_{\mathrm{ME}}^{(n)}(\ket{\psi})=\mathbb{E}[\pi_A] = \left(\begin{array}{l}n
\\n_A\end{array}\!\!\right)^{-1}\sum_{|A|=n_A}\pi_{A},
\end{equation}
where $\mathbb{E}$ denotes the expectation value, the combinatorial coefficient is the number of bipartitions, $|A|$ is the
cardinality of $A$ and the sum is over balanced bipartitions
$n_A=[n/2]$, where $[\cdot]$ denotes the integer part. The quantity $\pi_{\mathrm{ME}}$
measures the average bipartite entanglement over all possible
balanced bipartitions and inherits the bounds (\ref{eq:piArange})
\begin{equation}
\label{eq:pmeconstraint}
\pi_{\mathrm{min}}^{[n/2]} \le\pi_{\mathrm{ME}}^{(n)}(\ket{\psi})\le 1.
\end{equation}
A \emph{maximally multipartite entangled state} (MMES)
\cite{MMES} $\ket{\varphi}$ is a minimizer of
$\pi_{\mathrm{ME}}$,
\begin{eqnarray}
\label{eq:minimizer}
\pi_{\mathrm{ME}}^{(n)}(\ket{\varphi})= E_0^{(n)},  \\
\mathrm{with} \quad E_0^{(n)}=
\min \{\pi_{\mathrm{ME}}^{(n)}(\ket{\psi})\; | \; \ket{\psi}\in \mathcal{H}_S, \bra{\psi}\psi\rangle=1\}.
\nonumber\end{eqnarray}
The meaning of this definition is clear: most measures of bipartite entanglement
(for pure states) exploit the fact that when a pure quantum state
is entangled, its constituents are in a mixed state. We are simply
generalizing the above distinctive trait to the case of multipartite
entanglement, by requiring that this feature be valid for all
bipartitions. The density
matrix of each subsystem $A\subset S$ of a MMES is as mixed as possible (given
the constraint that the total system is in a pure state), so that
the information contained in a MMES is as distributed as possible.
The average purity introduced in Eq.\ (\ref{eq:piave}) is related to
the average linear entropy \cite{scott} and extends ideas put forward in
\cite{multipart4,parthasarathy}.

We shall say that a MMES is \emph{perfect} when the lower bound 
(\ref{eq:pmeconstraint}) is saturated 
\begin{equation}
\label{eq:perfectmmes}
E_0^{(n)} = \min \{\pi_{\mathrm{ME}}^{(n)}\} =\pi_{\mathrm{min}}^{[n/2]} . \nonumber
\end{equation}
It is immediate to see that a necessary and sufficient condition for a
state to be a MMES is to be maximally entangled with respect
to balanced bipartitions, i.e.\ those with $n_A=[n/2]$.
Since this is a very strong requirement, perfect MMES may not exist for $n>2$  (when $n=2$ the above equation can be
trivially satisfied) and the set of perfect MMES can be empty. 

In the best of all possible worlds one can still seek for  the
(nonempty) class of states that better approximate perfect MMESs,
that is states with minimal average purity. 
We shall say that a MMES is \emph{uniformly optimal} when its distribution of entanglement is as fair as possible, namely when the variance vanishes:
\begin{equation}
\label{eq:sigmaave}
\sigma^{(n)}
=\mathbb{E}\left[\left(\pi_A-\pi_{\mathrm{ME}}^{(n)}\right)^2\right]^{1/2} = 0,
\end{equation}
where the expectation is taken according to the same distribution as in Eq.\ (\ref{eq:piave}) (all balanced bipartitions). 
Of course, a perfect MMES is optimal. It is not obvious that uniformly optimal non-perfect MMES exist.

The very fact that perfect MMES may not exist is a symptom of frustration. We emphasize
that this frustration is a consequence of the conflicting requirements that entanglement be maximal for all possible bipartitions of the system.

\subsection{Qubits and the symptoms of frustration}
\label{sec:qubits}

For qubits the total Hilbert space is $\mathcal{H}_S= (\mathbb{C}^2)^{\otimes n}$ and factorizes into
$\mathcal{H}_S=\mathcal{H}_A\otimes\mathcal{H}_{\bar{A}}$, with
$\mathcal{H}_A= (\mathbb{C}^2)^{\otimes n_A}$, of dimensions
$N_A=2^{n_A}$ and $N_{\bar{A}}=2^{n_{\bar{A}}}$, respectively
($N_AN_{\bar{A}}=N$). 
Equations (\ref{eq:piArange}) and (\ref{eq:pmeconstraint})-(\ref{eq:minimizer}) read
\barr
\label{eq:purityconstraint}
 &  & 1/N_A\le\pi_{A}\le 1, \\
\label{eq:pmeconstraintQ}
& & 1/N_A \le E^{(n)}_{0}\le\pi_{\mathrm{ME}}^{(n)}(\ket{\psi})\le 1,
\qquad N_A= 2^{[n/2]},
\earr
respectively.

For small values of $n$ one can tackle the minimization problem
(\ref{eq:minimizer}) both analytically and numerically. For
$n=2,3,5,6$ the average purity saturates its minimum in
(\ref{eq:pmeconstraintQ}): this means that purity is minimal
\emph{for all} balanced bipartitions. In this case the MMES is \emph{perfect}.

For $n=2$ (perfect) MMES are Bell states up to local unitary
transformations, while for $n=3$ they are equivalent to the GHZ
states \cite{GHZ}. For $n=4$ one numerically obtains $E_0^{(4)}=\min
\pi_{\mathrm{ME}}^{(4)}=1/3 > 1/4 =1 /N_A$
\cite{MMES,sudbery,sudbery2,higuchi}. For $n=5$ and 6 one can find
several examples of perfect MESS  \cite{MMES,multent}.
The case $n=7$ is still open, our best estimate being $E_0^{(7)}
\simeq 0.13387 > 1/8 = 1 /N_A$. Most interestingly, perfect MMES  do
not exist for $n\geq 8$ \cite{scott}.
These findings are summarized in Table \ref{tab_qGmmes} (left column) and bring
to light the intriguing feature of multipartite entanglement we are interested in: since
the minimum $1/N_A$ in Eq.\ (\ref{eq:pmeconstraintQ}) cannot be
saturated, the value of $\pi_A$ must be larger for some bipartitions
$A$. We view this ``competition" among different bipartitions as a
phenomenon of frustration: it is already present for $n$ as small as
4. This frustration is the main reason for the difficulties one
encounters in minimizing $\pi_{\mathrm{ME}}$ in (\ref{eq:piave}).
Notice that the dimension of $\mathcal{H}_S$ is $N=2^n$ and the
number of partitions scales like $N$. We therefore need to define
a viable strategy for the characterization of the frustration in MMES, even for relatively small values of $n$.

\begin{table}[t]
\caption{Comparison between qubit and Gaussian maximally multipartite entangled states for different number $n$ of subsystems.}
\label{tab_qGmmes}
\begin{center}
\begin{tabular}{|c|c|c|}
\hline
$n$ & qubit perfect MMES & Gaussian perfect MMES \\
\hline
    2,3 & yes  & yes \\
    4 & no & no  \\
     5,6 & yes & no, but uniformly optimal$^*$\\
     7 & $\;\,$no$^*$ & no \\
    $\geq 8$ & no & no  \\
    \hline
\end{tabular}
\end{center}
\hspace*{6.2cm} $^*$numerical evidence
\end{table}

\subsection{Continuous variables and further symptoms of frustration}
\label{sec:contvar}

For continuous variables we have
$\mathcal{H}_S= (L^2(\mathbb{R}))^{\otimes n}$ with 
$\dim
L^2(\mathbb{R})=\infty$. As a consequence, the lower bound $\pi_{\mathrm{min}}^{n_A}=0$ in Eq.\ (\ref{eq:piallArange}) is not attained by any state. Therefore, strictly speaking, in this
situation  there do not even exist maximally \emph{bipartite} entangled states,
but only states that approximate them. This inconvenience can be
overcome by introducing physical constraints related to the limited
amount of resources that one has in real life. This reduces the set
of possible states and induces one to reformulate the question in
the form: what are the physical minimizers of (\ref{eq:piAdef}),
namely the states that minimize (\ref{eq:piAdef}) and belong to the
set $\mathcal{C}$ of physically constrained states? In sensible
situations, e.g.\ when one considers states with bounded energy and
bounded number of particles,  the purity lower bound
\begin{equation}
\pi_{\mathrm{min}}^{n_A, \mathcal{C}} = \inf\{\pi_A, |\psi\rangle\in\mathcal{C} \}\geq \pi_{\mathrm{min}}^{n_A}
\end{equation}
is no longer zero and is attained by a class of minimizers, namely
the \emph{maximally bipartite entangled states}. If this is the
case, we can also consider multipartite entanglement and ask whether
there exist states in $\mathcal{C}$ that are maximally entangled for
every bipartition $(A,\bar{A})$, and therefore satisfy the extremal
property
\begin{equation}
\pi_A = \pi_{\mathrm{min}}^{n_A, \mathcal{C}}
\label{eq:perfectMMESdef}
\end{equation}
for every subsystem $A\subset S$ with $n_A=|A| \leq n/2$. In analogy
with the discrete variable situation, where $\dim
\mathfrak{h}<\infty$ and $\mathcal{C}=\mathcal{H}$, we will call a
state that satisfies (\ref{eq:perfectMMESdef}) a \emph{perfect
MMES} (subordinate to the constraint $\mathcal{C}$).

Since, once again, the requirement (\ref{eq:perfectMMESdef}) is very strong, the
answer to this quest can be negative for  $n>2$  (again, when $n=2$ it is
trivially satisfied) and the set of perfect MMES can be empty. In the best of the best of all possible worlds one can still seek for  the (nonempty) class of states that better approximate perfect MMESs,
that is states with minimal average purity. 
In conclusion, by definition a MMES is a state that  belongs to $\mathcal{C}$ and
minimizes the potential of multipartite entanglement (\ref{eq:piave}). Obviously,
when 
\begin{equation}
\label{eq:perfectmmesC}
E_0^{(n),\mathcal{C}} = \min_{\mathcal{C}}  \{\pi_{\mathrm{ME}}^{(n)}\} =\pi_{\mathrm{min}}^{[n/2],
\mathcal{C}} 
\end{equation}
there is no frustration and the MMESs are perfect.
Eventually, we will consider the limit $\mathcal{C}\rightarrow\mathcal{H}$.

Let us consider the quantum state $|\psi_{(n)}\rangle$ of $n$  identical bosonic 
oscillators with (adimensional) canonical variables $\{ q_k,
p_k \}_{k=1,\dots n}$ and unit frequency (set $\hbar=1$). An analogous description of the system can be given in terms of 
the Wigner function on the $n$-mode phase space
\begin{equation}\label{Wigner}
W_{(n)}(q, p) = \int d^n{y} \langle
q -  y | \psi_{(n)}\rangle \langle\psi_{(n)} |  q +  y
\rangle e^{2 i \pi y\cdot p},
\end{equation}
where $ q = (q_1, \dots q_n) $, $ p = (p_1, \dots
p_n) $, $ y = (y_1, \dots y_n) \in\mathbb{R}^n$, and we
have denoted by
\begin{equation}
|  q \pm  y \rangle = \otimes_{k=1}^n | q_k \pm y_k
\rangle
\end{equation}
the generalized position eigenstates. By definition, Gaussian states \cite{cvbooks,cvbooks1} are those described by a Gaussian
Wigner function. Introducing the phase-space coordinate vector 
$ X = (X_1, \dots X_{2n}) = (q_1, p_1, \dots q_n, p_n)$, a
Gaussian state has a Wigner function of the following form:
\begin{eqnarray}
W_{(n)}(X) &=& \frac{1}{(2\pi)^n\sqrt{\det(\mathbb{V})}}\nonumber\\
&\times&\exp{\left[-\frac{1}{2}(X- X_0)
\mathbb{V}^{-1}(X- X_0)^\mathsf{T}\right]},
\end{eqnarray}
where $X_0 = \langle X \rangle= \int X W_{(n)}(X)
d^{2n} X$, is the vector of first moments, and $\mathbb{V}$
is the $2n\times 2n$ covariance matrix, whose elements
are
\begin{equation}
\mathbb{V}_{lm} = \langle ( X_l - \langle X_l \rangle )( X_m -
\langle X_m \rangle ) \rangle.
\end{equation}
For Gaussian states, purity is a function of the
``sub"determinant of the covariance matrix 
\begin{equation}\label{purity}
\pi_A = \frac{1}{2^{n_A}\sqrt{\det (\mathbb{V_A})}},
\end{equation}
$\mathbb{V_A}$ being the square submatrix defined by the indices pertaining to bypartition $A$.
Clearly, $\pi_A  \le 1$.

The results of the search for perfect and uniformly optimal MMES with Gaussian states are summarized in the right column of Table \ref{tab_qGmmes} \cite{FFLMP}. There are curious analogies and differences with qubit MMES. In particular, for $n\ge 8$ perfect MMES do not exist in both scenarios.  Actually, for Gaussian states, frustration is present for $n\ge 4$; for the ``special" integers $n=5,6$ we notice that both for two-level and continuous variables systems the variance of the distribution of entanglement goes to zero; on the other hand, in the former case one can find perfect MMES, in the latter case MMES are uniformly optimal but not perfect.

\section{Scrutinizing Frustration}
\label{sec:frustr7}

We now turn to the detailed study of the structure of frustration. 
For the sake of concreteness, we shall first focus on qubits.
Let us start from a few
preliminary remarks. We observed that it is always possible to saturate the lower
bound in (\ref{eq:purityconstraint})
\begin{equation}
\label{eq:pmalb}
\pi_{A}= 1/N_A
\end{equation}
for a \emph{given} balanced bipartition $(A,\bar A)$ with $N_A=2^{[n/2]}$. However, in order to
saturate the lower bound in (\ref{eq:pmeconstraintQ})
\begin{equation}
\label{eq:pmelb}
E_0^{(n)}=1/N_A,
\end{equation}
condition (\ref{eq:pmalb}) must be valid for every bipartition in the average
(\ref{eq:piave}). As we mentioned in Sec.\ \ref{sec:qubits}, this requirement can be satisfied
only for very few ``special" values of $n$ ($n=2,3,5$ and 6, see
Table \ref{tab_qGmmes}). For all other values of $n$ this is
impossible: different bipartitions ``compete" with each other, and
the minimum $E_0^{(n)}$ of $\pi_{\mathrm{ME}}^{(n)}$ is strictly
larger than $1/N_A$.

\begin{figure}
\begin{center}
\includegraphics[width=0.7\textwidth]{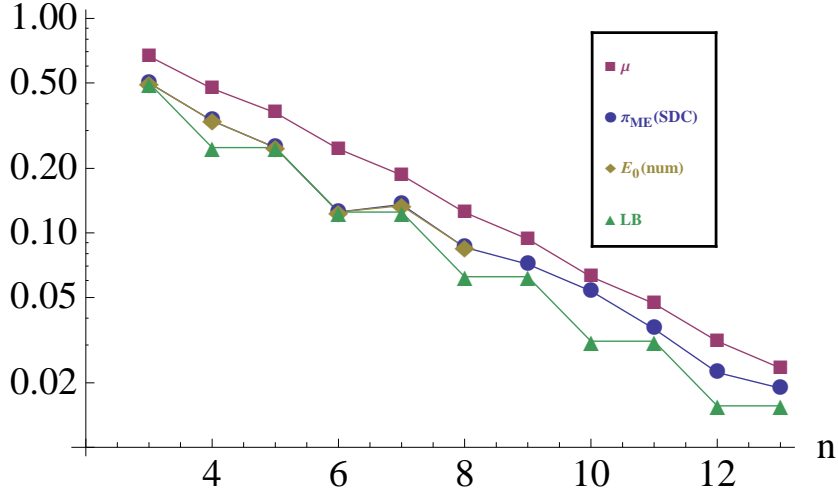}
\caption{(Color online)
Qubits: average purity $\mu$ of the typical states (squares), average purity
$\pi_{\mathrm{ME}}(\mathrm{SDC})$ computed according to Scott's
procedure \cite{scott} with extremal additive self-dual codes (full
circles), our best numerical estimate $E_0(\mathrm{num})$ for the
minimum of $\pi_{\mathrm{ME}}$ (diamonds) and lower bound
$\mathrm{LB}=1/N_A$ (triangles) vs the number of spins $n$. Note
that for $n=7$ the full circle and the diamond do not coincide. The
scale on the ordinates is logarithmic.}
\label{fig55}
\end{center}
\end{figure}

It is interesting to look at this phenomenon in more detail. 
Let us recall that for typical states
\cite{lubkinrnd,lloydpagelsrnd,pagernd,zyczkowskirnd,scottcavesrnd,FFP,giraudrnd}
the distribution of purity over balanced bipartitions has mean
 \beq
\mu^{(n)} = \frac{N_A+N_{\bar{A}}}{N+1}. \label{eq:6a} 
\eeq
Figure \ref{fig55} displays the average purity $\mu$ of typical states [Eq.\ (\ref{eq:6a})], the average purity of extremal additive self-dual
codes states, computed according to Scott's procedure \cite{scott},
our best numerical estimate for the minimum of $E_0^{(n)}$, and the
lower bound $1/N_A$. All these quantities exponentially vanish as $n
\to \infty$. Scott's states give an upper bound for the minimal average purity when $n\geq 8$, where the numerical simulations become very time consuming. In particular, we notice that for $3 \le n \le 6$ and $n=8$ the numerical values of $E_0$ coincide with the results obtained using extremal
additive self-dual codes \cite{scott}. For $n=7$ the optimization algorithm reaches a lower value. For $n>8$ our numerical data do not enable us to draw any conclusions.

\begin{figure}
\begin{center}
\includegraphics[width=0.7\textwidth]{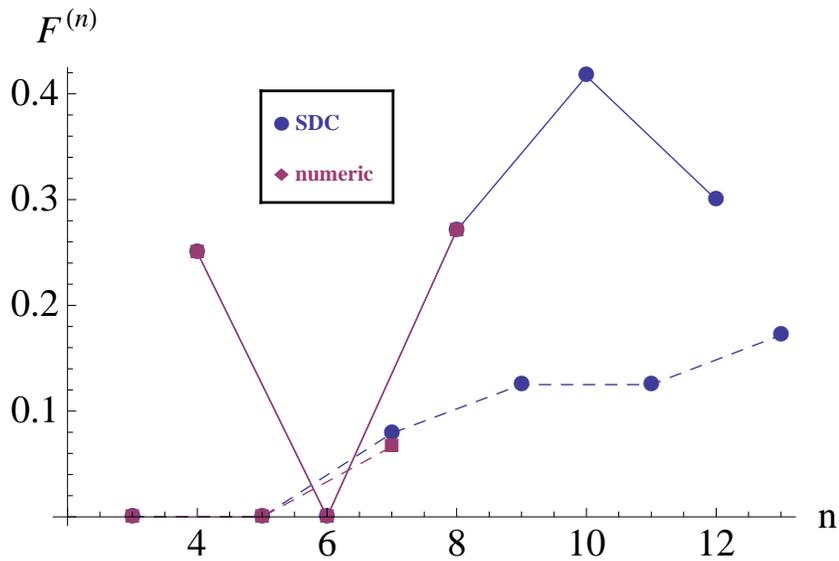}
\caption{(Color online)
Qubits: frustration ratio: (normalized) distance between the average purity
$\pi_{\mathrm{ME}}(\mathrm{SDC})$, according to Scott's extremal
additive self-dual codes (full circles), or our best numerical
estimate for the minimum $E_0(\mathrm{num})$ (squares), and the
lower bound LB=$1/N_A$, vs the number of spins. Notice the
difference between even and odd $n$. }
\label{fig66}
\end{center}
\end{figure}

Figure \ref{fig66} displays the (normalized) difference between the
minimum average purity, computed according to Scott's extremal
additive self-dual codes and/or our best numerical
estimate, and the lower bound $1/N_A$ in Eq.\ (\ref{eq:pmelb}). This
difference is an upper bound to the \emph{frustration ratio}
\begin{equation}
\label{eq:Frn}
F^{(n)}= \frac{E_0^{(n)}-1/N_A}{E_0^{(n)}}, \qquad N_A= 2^{[n/2]} ,
\end{equation}
that can be viewed as the ``amount of frustration" in the system. 
We notice the very different behavior between odd and even values of $n$. In the former case the amount of frustration increases with the size of the system. On the contrary, in the latter case, the behavior is not monotonic. It
would be of great interest to understand how this quantity behaves
in the thermodynamical limit, but our data do not enable us to draw
any clear-cut conclusions. 

\begin{figure}
\centering
\includegraphics[width=0.5\textwidth]{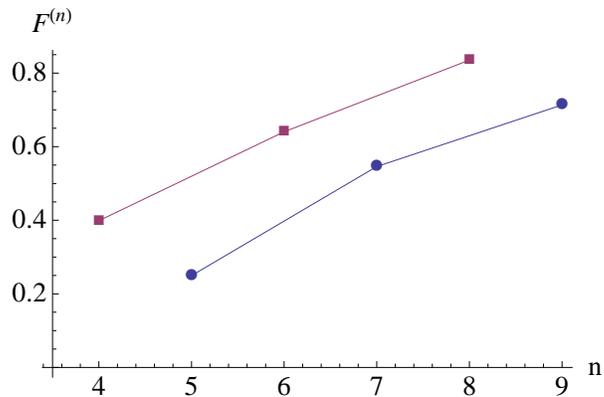}
\caption{(Color online) Gaussian states: frustration ratio for and odd (rectangle) and even (circle) number of modes. As explained in the text, each point has been obtained in the saturation regime $\mathcal{C}\to\mathcal{H}$.} 
\label{gaussfrustplot}
\end{figure}

Let us extend these considerations to the continuous variables scenario. 
In order to measure the amount of frustration (for states belonging to the constrained set $\mathcal{C}$) we define a more general frustration ratio
\begin{equation}
\label{merit1}
F^{(n),\mathcal{C}}= \frac{E_0^{(n),\mathcal{C}}-\pi_{\mathrm{min}}^{[n/2],
\mathcal{C}} }{E_0^ {(n),\mathcal{C}}}
\end{equation}
and eventually take the limit $\mathcal{C}\rightarrow\mathcal{H}$, where both the numerator denominator can vanish. [Notice that (\ref{eq:Frn}) is a specialization of the quantity in
(\ref{merit1}) to the qubit case, i.e.\ $\mathcal{C}=\mathcal{H}= (\mathbb{C}^2)^{\otimes n}$.] As a constraint we fix the value $\mathcal{N}$ of the average number of excitations per mode, namely
\begin{equation}
\label{eq:constraint}
\mathcal{C}=\{\psi\in\mathcal{H},  \psi\; \mathrm{Gaussian}, 
\frac{\langle q_k^2 + p_k^2 \rangle}{2} \le \mathcal{N} + \frac{1}{2},  1\leq k\leq n\} .
\end{equation}
The ideal lower bound  is given by \cite{FFLMP}
\begin{equation}\label{min_purity}
 \pi_{\mathrm{min}}^{[n/2],\mathcal{C}}=
\frac{1}{2^{[n/2]}(\mathcal{N}+1/2)^{[n/2]}}
\end{equation}
and represents the purity of a Gaussian thermal state. Incidentally, we notice that, for $\mathcal{N}=1/2$, Eq.\ (\ref{min_purity}) reproduces the lower bound of Eq.\ (\ref{eq:pmeconstraintQ}) i.\ e.\ the case of qubits. On the other hand, $\mathcal{N}=0$ corresponds to the case of completely separable states.
In Fig.\  \ref{gaussfrustplot} we plot the frustration ratio (\ref{merit1}) as a function of the number of modes. Each point has been numerically obtained by relaxing the energy constraint ($\mathcal{N}\rightarrow +\infty$) until the ratio $F^{(n),\mathcal{C}}$ has reached a saturation value $F^{(n)}$ \cite{FFLMP}. This corresponds to the limit $\mathcal{C}\to\mathcal{H}$.

\subsection{The structure of frustration}
\label{sec:scrtuctfrustr}

\begin{figure}
\begin{center}
\includegraphics[width=0.7\textwidth]{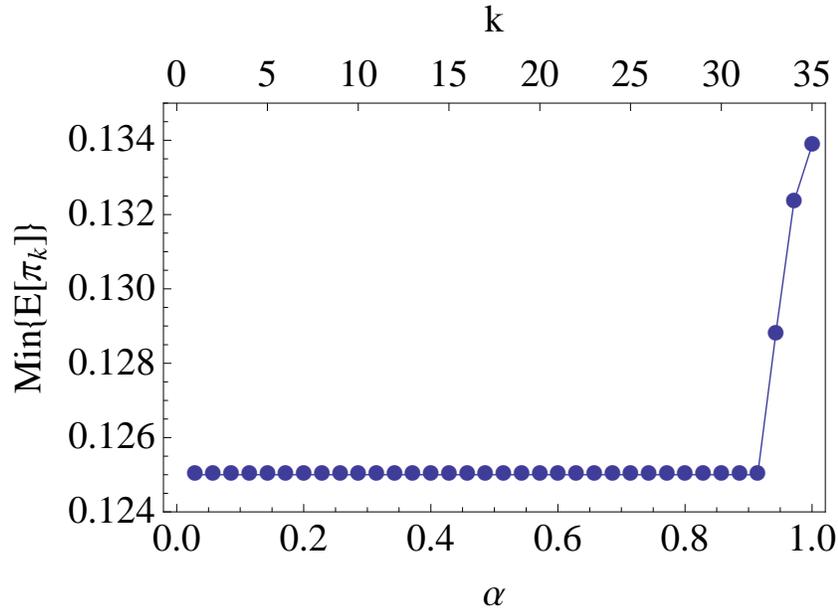}
\caption{(Color online)
Minimum average purity for $n=7$ and an increasing number $k$ of
bipartitions. Until $k=32$ no frustration appears. For $k=33$ one
partition has purity $\pi=1/4$. For $k=34$, two bipartitions yield
$\pi=1/4$. For $k=K=35$, $\pi > 1/8$ for all bipartitions. $\alpha=k/K$. See Fig.\
\ref{figbip}.}
\label{figf7}
\end{center}
\end{figure}

\begin{figure}
\begin{center}
\includegraphics[width=0.9\textwidth]{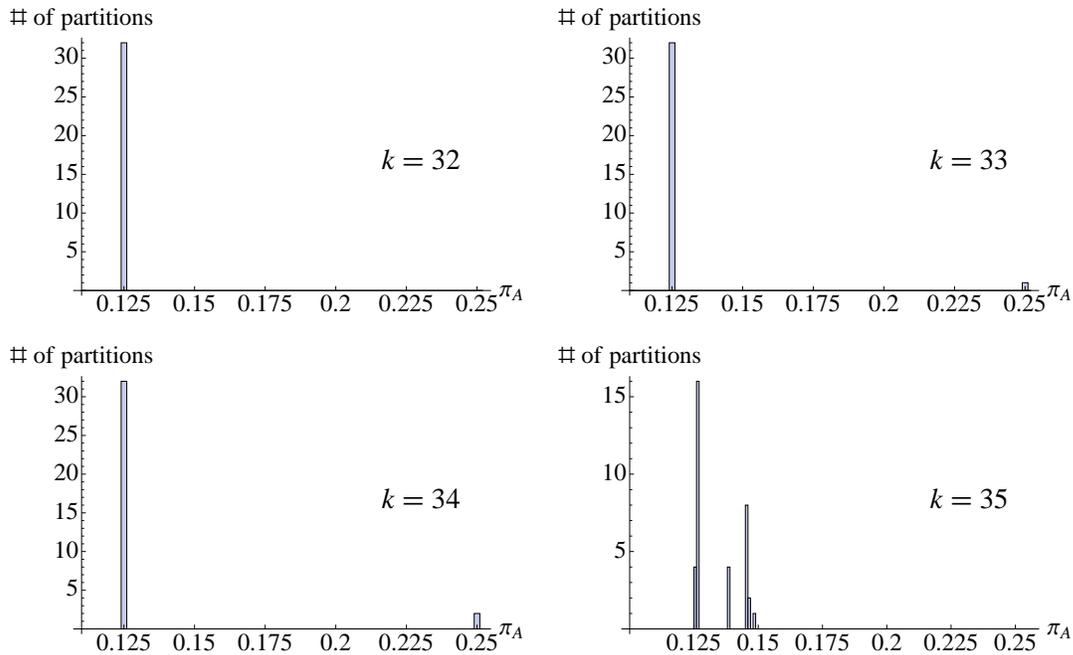}
\caption{(Color online)
Distribution of purity for $n=7$ and an increasing number $k$ of
bipartitions. See Fig.\ \ref{figf7}. }
\label{figbip}
\end{center}
\end{figure}

In order to try and understand the underlying structure of this frustration, we focus again on qubits and
analyze the behavior of the minimum value of
the average purity when one requires the condition (\ref{eq:pmalb})
for an increasing number of bipartitions. We proceed as follows: we set $n=7$, so that the
total number of balanced bipartitions is $K=35$, and add bipartitions one by one, by choosing them so that 
condition (\ref{eq:pmalb}) be valid, as far as this is possible. When (\ref{eq:pmalb}) becomes impossible to satisfy, we require that the average purity be minimal.
We plot the minimum average purity as a function of $k$ and $\alpha=k/K$ in Fig.\ \ref{figf7} .
One observes that
it is possible to saturate the minimum $1/N_A=1/8=0.125$ up to
$k=32$ well chosen partitions. For $k=33$ all bipartitions yield the minimum 1/8, except
the last one, that yields 1/4: frustration appears. For $k=34$, two
bipartitions yield 1/4. For $k=K=35$ purity is larger than 1/8 for \emph{all}
bipartitions. Notice that the solution with 32 bipartitions at 1/8
and the remaining \emph{three} at 1/4 corresponds to Scott's extremal
additive self-dual code and would yield a higher average. The
distribution of purity for an increasing number of partitions is
shown in Fig.\ \ref{figbip}.

\begin{figure}
\begin{center}
\includegraphics[width=0.7\textwidth]{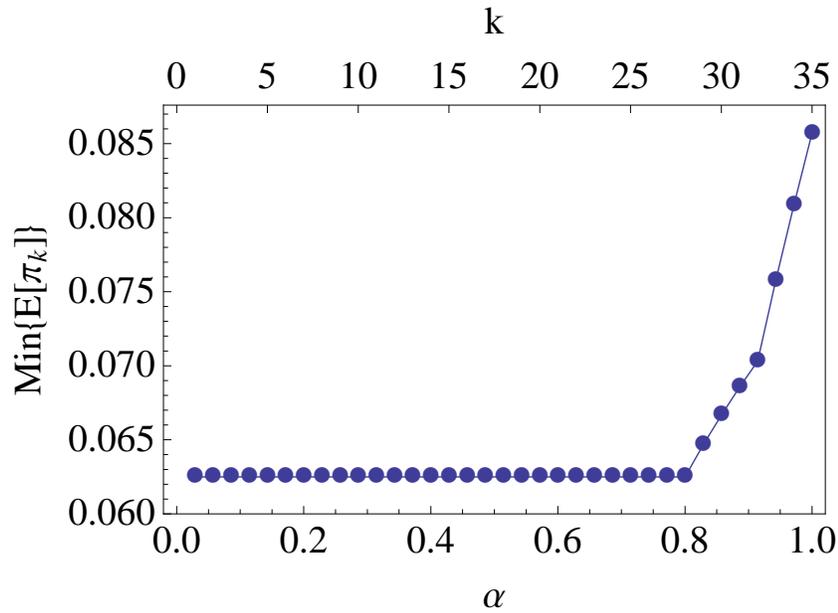}
\caption{(Color online) 
Minimum average purity for $n=8$ and an increasing number $k$ of
bipartitions. Until $k=28$ no frustration appears.  $\alpha=k/K$.}
\label{figbip2}
\end{center}
\end{figure}

The case $n=8$ is slightly different: see Fig.\ \ref{figbip2}. In this case $K=35$ again and there is no frustration up to $k=28$ bipartitions ($\alpha=0.8$), if properly chosen (all of them with a purity $1/16$). When $k$ is further increased, it is no longer possible to reach the lower bound: for $29\le k \le 32$ the new bipartitions have purity $1/8$. Finally, for $33\le k \le 35=K$, the new bipartitions have purity $1/4$.

\begin{figure}
\begin{center}
\includegraphics[width=0.7\textwidth]{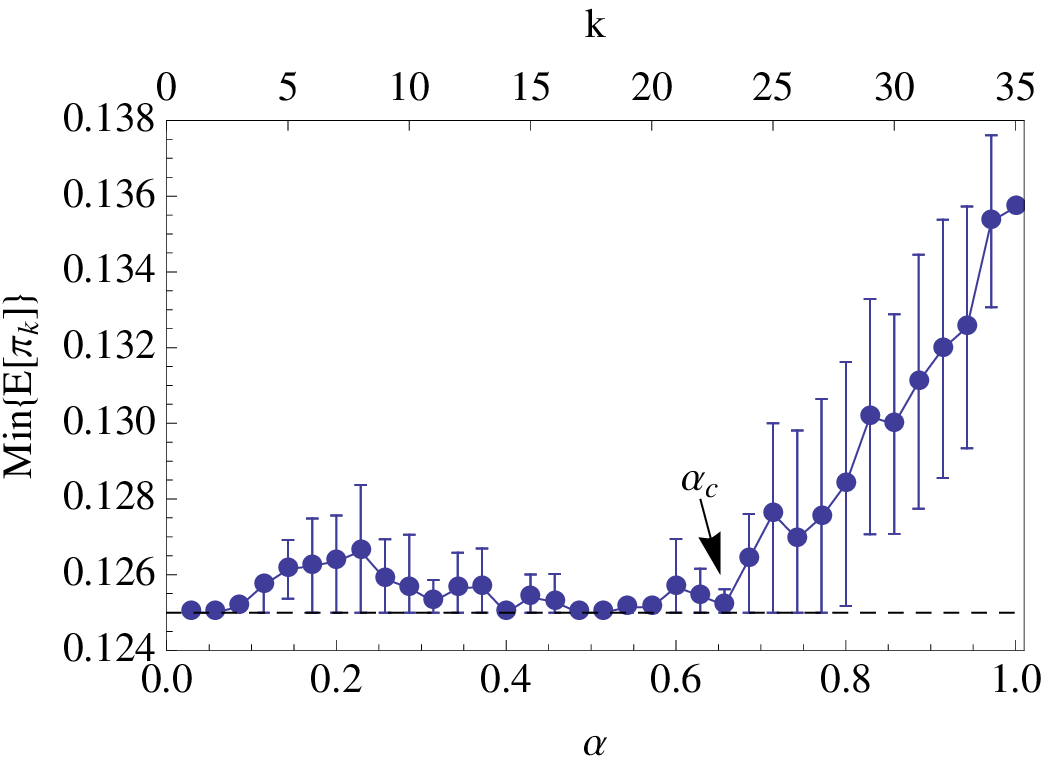}
\caption{(Color online)
Minimum average purity for $n=7$ and an increasing number $k$ of randomly selected
bipartitions. Every point corresponds to an average over different extractions of $k$ bipartitions. The bars correspond to 68\% of the distribution (that can be very asymmetric due to the closeness of the lower bound). 
 $\alpha=k/K$. The average fraction of frustrated bipartitions is $1-\alpha_c \simeq 34\%$.}
\label{figbip3}
\end{center}
\end{figure}

Another useful test is the extraction of $k$ \emph{randomly} selected bipartitions and the successive minimization of the average purity. This is a typical test in frustrated systems, e.g.\ in random \cite{random} and Bethe lattices \cite{bethe}. In this way one checks the onset of frustration independently of the particular choice of the sequence of bipartitions. In Fig.\ \ref{figbip3} we plot the dependence on $k$ and $\alpha$. Each point corresponds to the extraction of a number of $k$ bipartitions ranging from few tens to a few hundreds. Frustration appears at rather large values of $k$, qualitatively confirming the result shown in Fig.\ \ref{figf7}. Moreover we notice that for smaller $k$ it is sometimes difficult to reach the minimum. This could be an indicator that for a small number of bipartitions there is a large number of local minima in the energy landscape, that ``traps" the numerical procedure.
Notice that the curve in Fig.\  \ref{figbip3} should monotonically increase as a function of $k$, so that all deviations from monotonicity are ascribable to the numerical procedure, and are a consequence of the fact that for different values of $k$, in each run of the simulation, the subset of extracted bipartitions is uncorrelated to the set used for the preceding values of $k$.

Although the results of this section are not conclusive, they provide a
clear picture of the relationship between entanglement and
frustration. The latter tends to grow with the size of the system
(Figs.\ \ref{fig66} and \ref{gaussfrustplot}) and it is difficult to study, at least for small
values of $n$, because it suddenly appears at the last few
bipartitions (Fig.\ \ref{figf7}). One estimates, from the results for $n=7$ qubits in Fig.\ \ref{figbip3}, an average fraction of frustrated bipartitions $1-\alpha_c \simeq 34\%$, $\alpha_c$ being a critical ratio.

A posteriori, it is not surprising
that multipartite entanglement, being a complex phenomenon, exhibits
frustration. It would be of great interest to understand what
happens for larger values of $n$.
Different scenarios are possible, according to the mutual interplay between the quantities $F$ and $\alpha_c$. In particular, the amount of frustration $F$ shows a tendency to increase with $n$, for both qubit and Gaussian states. This could be ascribable to a decrease of $\alpha_c$, corresponding to an increasing fraction of frustrated partitions, or to a constant (or even increasing) $\alpha_c$, corresponding to a constant (or decreasing) fraction of increasingly frustrated bipartitions.

\section{Concluding remarks}\label{sec:conclusion}
\label{sec:concl}
One important property that we have not investigated here and that
is often used to characterize multipartite entanglement is the
so-called monogamy of entanglement \cite{multipart1,KDS}, that
essentially states that  entanglement cannot be freely shared among
the parties. Interestingly, although monogamy is a typical property
of multipartite entanglement, it is expressed in terms of a bound on
a sum of \emph{bipartite} entanglement measures. This is reminiscent
of the approach taken in this paper. The curious fact that bipartite
sharing of entanglement is bounded might have interesting
consequences in the present context. It would be worth understanding
whether monogamy of entanglement generates frustration.

Two crucial issues must be elucidated. First, the striking similarities and small differences between qubits and Gaussian MMES: see Table \ref{tab_qGmmes} and compare Figs.\  \ref{fig66} and \ref{gaussfrustplot}. Second, the features of MMES for $n\to \infty$. Finally, we think that the characterization of multipartite entanglement investigated here can be important for the analysis of the entanglement features of
many-body systems, such as spin systems and systems close to criticality.

\ack
We thank C.\ Lupo, S.\ Mancini and A.\ Scardicchio for interesting discussions. This work is partly supported by the European Community
through the Integrated Project EuroSQIP.

\end{document}